\documentclass[pra,showpacs,twocolumn,superscriptaddress,aps]{revtex4-1}%
\usepackage{color}
\usepackage{amsfonts}
\usepackage{amsmath}
\usepackage{amssymb}
\usepackage{mathrsfs}
\usepackage{graphicx}%
\usepackage{bm}
\usepackage{multirow}
\usepackage{epstopdf}

\begin{document}

\title{Quantum feedback control of atomic ensembles and spinor Bose-Einstein condensates}

\author{Shi Wang}
\affiliation{University of New South Wales, Canberra, Australian Capital Territory 2600, Australia}
\affiliation{National Institute of Informatics, 2-1-2 Hitotsubashi, Chiyoda-ku, Tokyo 101-8430, Japan}

\author{Tim Byrnes}
\affiliation{New York University, 1555 Century Ave, Pudong, Shanghai 200122, China}
\affiliation{NYU-ECNU Institute of Physics at NYU Shanghai, 3663 Zhongshan Road North, Shanghai 200062, China}
\affiliation{National Institute of Informatics, 2-1-2 Hitotsubashi, Chiyoda-ku, Tokyo 101-8430, Japan}

\begin{abstract}
Cold atomic ensembles and spinor Bose-Einstein condensates (BECs) are potential candidates for quantum memories as they have long coherence times and can be coherently controlled.  Unlike most candidates for quantum memories which are genuine or effective single particle systems, in atomic ensembles the quantum information is stored as a spin coherent state involving a very large number of atoms. A typical task with such ensembles is to drive the state towards a particular quantum state. While such quantum control methods are well-developed for qubit systems, it is a non-trivial task to extend quantum control methods to the many-particle case.  The objective of this work is to deterministically steer an arbitrary state of the atomic ensemble into a desired spin coherent state. To this end, we design our control law using stochastic stability theory, the quantum filtering theorem, and phase contrast imaging.  We apply our control laws to different axes and show that it is possible to manipulate the atoms into different target states.
\end{abstract}


\maketitle

\section{Introduction}
\label{sec:intro}

Recent advances in the experimental techniques of cold atomic ensembles and Bose-Einstein condensates (BECs) have sparked the interests of researchers in their application toward quantum information tasks, in particular quantum metrology \cite{treutlein2006,bohi2009,riedel2010,gross2012} and quantum simulation \cite{buluta2009,bloch2005}.  For quantum computing, the traditional view is that single (or effectively single) particle systems are preferable, with the leading candidates being qubits made with ion trap, superconducting, N-V center, quantum dot, and photon technologies \cite{ladd2010}.  Nevertheless quantum memories based on ensembles, rather than single particles, remain attractive due to both experimental advantages and fundamental differences to qubit approaches.  Several approaches have been proposed for the use of ensembles as quantum memories in quantum information processing applications.  The first is a collective state encoding where excited states involving all the particles in the ensemble are used \cite{brion2007,lukin2001}. In such schemes discrete states are used to encode the quantum information, and has been used to perform fundamental tasks such as quantum teleportation \cite{bao2012}.  The second approach is to use the ensembles to approximate continuous variable quadrature variables \cite{julsgaard2001,krauter2012,braunstein2005}. In this approach only states that are in the vicinity of a particular total spin polarization direction (usually taken to be $ S^x $) are used, and the remaining total spins are used as the quadratures ($S^y$ and $ S^z $ in this case).  Another approach is to take advantage of the similarity of the mathematical structure of spin coherent states to qubits, to perform quantum information processing much in the same way as qubits, but using ensembles \cite{byrnes2012,byrnes2014,pyrkov2013}.

In order to measure the state of atomic ensembles and BECs, various optical imaging techniques including absorption imaging \cite{anderson1995,andrews1997,vestergaardhau1998}, fluorescent imaging \cite{depue2000} and phase contrast imaging (PCI) \cite{bradley1997} have been developed. PCI is an example of a non-destructive technique which does not destroy the BEC itself during the measurement process.  It is also a weak measurement in the sense that the quantum state is approximately preserved after the measurement process \cite{ilookeke2014}, as the number of scattered photons in the measurement can be made negligible. This opens the opportunity to use such methods beyond measurements, and can be incorporated to perform quantum control of the ensemble. Such a scheme was experimentally demonstrated in Ref. \cite{vanderbruggen2013}, where atomic ensembles affected by collective noise were weakly measured optically, and corrected back to their original position on the Bloch sphere.  This allows for a way to fight decoherence for collective noise in such ensembles by continuously (weakly) measuring the ensemble. This technique is also of relevance to atomic clocks to reduce the frequency noise of a local oscillator \cite{shiga2012}.

\begin{figure}
\centering
\includegraphics[width=\columnwidth]{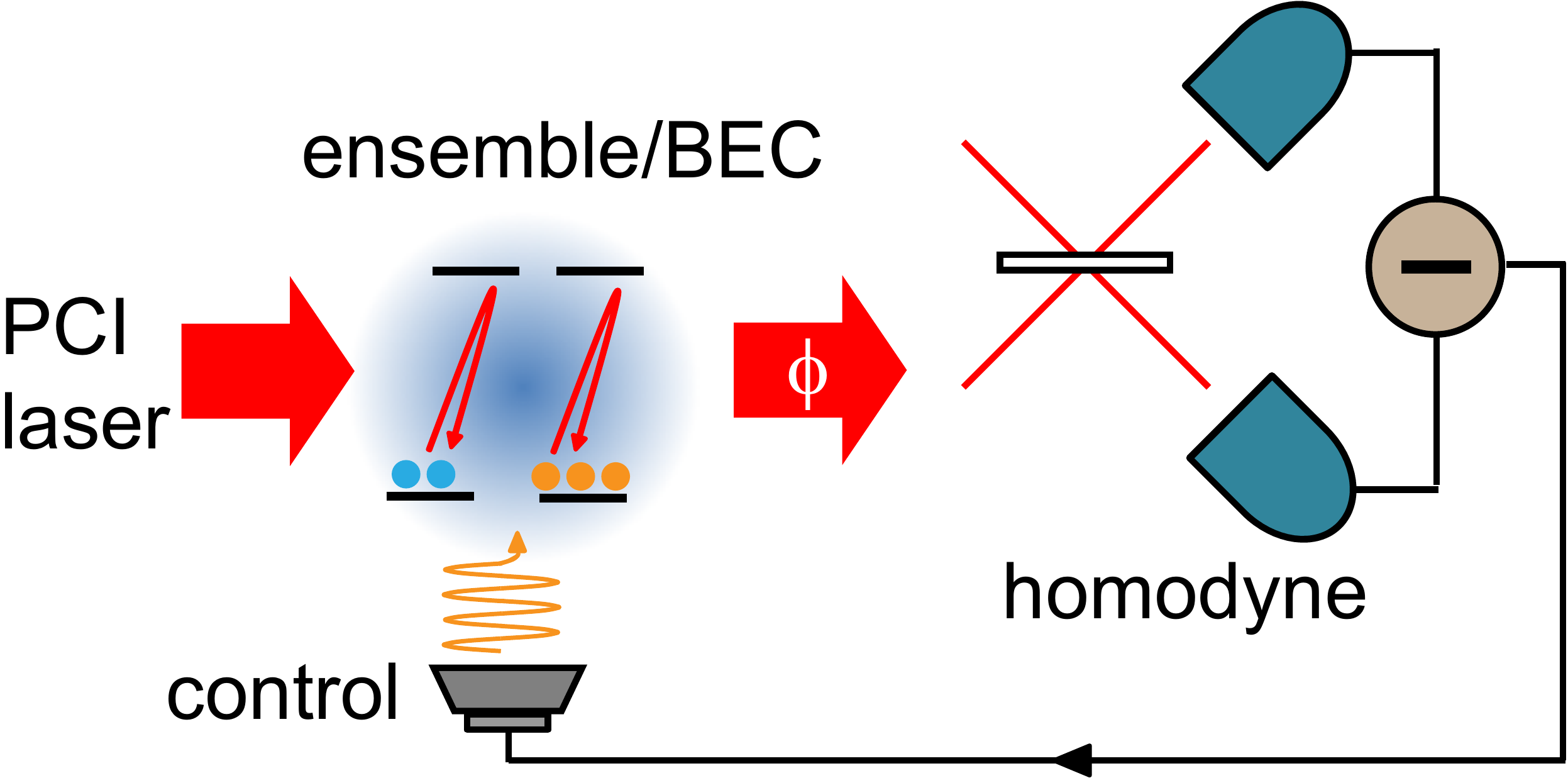}
\caption{\label{fig:bec} The phase contrast imaging (PCI) measurement based quantum feedback control scheme considered in this paper. The atom ensemble or BEC is illuminated with an ac Stark shift laser, which is affected witha phase shift depending on the state of the atoms. The phase shift is measured via a homodyne measurement and converted to an electrical signal.  The detected signal is processed by a
controller which feeds back to the BEC  to control the state of the atoms via a collective operation. }
\end{figure}

In the quantum feedback approach of Ref. \cite{vanderbruggen2013}, the aim is to stabilize one particular state by detecting the small deviations from the target state, and applying a rotation to counteract the original noise.  A more general task is to target an arbitrary state given an arbitrary initial state, where the deviation is not necessarily small. Such tasks are central to quantum control theory \cite{WM93,W94,DJ99,DHJMT2000,WD2005,WM2009}, and have been successfully applied to atomic and molecular physics as well as physical chemistry \cite{CHUS02,WHLRW13}. These have been mainly analyzed for systems with small Hilbert space dimensions such as qubits \cite{WB08,HKYDI2012}, however, in the above context it is an important task to extend this towards systems involving many particles. The purpose of this paper is to develop quantum control methods to the atomic ensembles and BECs, so that one can drive an arbitrary state towards a given target state. We note that there have been previous works that have also investigated quantum control problems relating to BECs \cite{szigeti2009,szigeti2010}.  These have been mainly focused on controlling the spatial modes of the BEC, and thereby driving the system into the ground state.  In our study we will instead consider rather the internal degrees of freedom of a two-component spinor BEC.  Whether the state can be effectively controlled is of high relevance to applications such as quantum information processing where the ensemble or BEC is used as a quantum memory.

More specifically, we will devise a quantum feedback scheme based continuous PCI measurements and control.  The measurement quantum feedback control scheme is shown in Fig. \ref{fig:bec}. The atom ensemble is illuminated by a laser field which induces an ac Stark shift upon the atoms.  For an ensemble with internal degrees of freedom, this is an entangling interaction, and the light carries information relating to the internal state of the atoms \cite{ilookeke2014}.  The light is interfered using homodyne detectors and the measurement result is compared with the desired value of the output.  Based on this control signal, the ensemble is controlled by a collective operation using a suitable control law. In quantum feedback control, the stochastic master equation (SME) plays an important role for the design of the feedback control system since it allows us to calculate both
the state of the system conditioned on a given set of measurement outcomes. This allows us to use
stochastic Lyapunov techniques and LaSalle's invariance principle to design a feedback control law \cite{LL1961,wang2010}.

This paper is organized as follows.  In Sec. \ref{sec:modeling} we describe our model of atomic ensembles and BECs, and derive the conditional master equation which forms the basis of the feedback theory. The main result in this section is Eq. \ref{conditional-master1} where the dynamics of the total spin is derived in the presence of feedback.   In Sec. \ref{sec:num} we derive the control laws for driving a particular target quantum state. This is derived from LaSalle's invariance principle where we argue the correct form of the feedback Hamiltonian.  For the disinterested reader who is only interested in the main results of this paper, one may start immediately at Sec. \ref{Simulation} where the conditional master equations are tested using numerical simulations.   We finally summarize our findings of this work in Sec. \ref{sec:conc}.

\section{Stochastic master equation}
\label{sec:modeling}

In this section we derive the stochastic master equation which describes the quantum feedback process for the PCI homodyne measurement.  We first derive the PCI interaction which gives rise to the estimations of feedback signals. This will serve to set up the approximations under which we work, leading us to the formalism of the total spin collective operations. Finally, the stochastic master equation which determines the effect of the whole system is derived.

\subsection{Phase contrast imaging Hamiltonian and BEC model}

In phase contrast imaging an off-resonant laser field illuminates an ensemble or BEC of atoms inducing an ac Stark shift (see Fig. \ref{fig:bec}).  The off-resonant detuning of the light induces a second order transition in the excited states of the atoms. A detailed theory of phase contrast imaging is given in Ref. \cite{ilookeke14}.  In this section  we derive the effective Hamiltonian, measurement operators, and conditional master equation specializing to the BEC case.

As with any derivation of a master equation, we divide the total Hilbert space into two parts, the degrees of freedom of interest (i.e. the ``system'') and the remaining parts (i.e. the ``environment''). The system in this case is the BEC, given by the Hamiltonian
\begin{align}
\hat{H}_{\mathrm{S}}=& \sum_j \int d\bm{x} \Big[\hat{\psi}_{gj}^\dagger(\bm{x})H_{nj} (\bm{x}) \hat{\psi}_{gj}(\bm{x}) +\nonumber\\
& \hat{\psi}_{ej}^\dagger(\bm{x}) H_{nj} (\bm{x}) \hat{\psi}_{ej} (\bm{x})\Big],
\end{align}
where  the field operator is $\hat{\psi}_{g j}(\emph{\textbf{x}})$ (or $\hat{\psi}_{e j}(\emph{\textbf{x}})$) annihilates an atom in the ground (or excited)  state with component $j$ at position $ \bm{x} $ (for the case of  two-component BECs,  the sum thus runs over $ j =1,2 $). The mean field single atom Hamiltonian is
\begin{align}
H_{nj} (\bm{x}) = \hbar \omega_{nj} -\frac{\hbar^2}{2m} \nabla^2 + V(\bm{x}) + \kappa_{nj} | \psi_0 (\bm{x})|^2,
\label{singleparticleham}
\end{align}
where $ \omega_{nj} $ is the frequency of the internal state labeled by $ n$ and $j $   ($n = g,e$),
$m$ is the mass of the atom, $ V(\bm{x}) $ is the trapping potential, $ \kappa_{nj} $ is the atomic interaction between atoms in the state $ n, j $ with the ground state atoms, and $ \psi_0 (\bm{x}) $ is the mean field wavefunction of the ground state atoms assumed to be invariant throughout the subsequent dynamics.

While we primarily consider only two components $ j = 1,2 $ for this paper, this can be straightforwardly extended to any number of components.  The two component case is most relevant for approaches where two hyperfine ground states are used as the storage states.  For magnetically trapped BECs, only magnetic sublevels with the correct parity can be trapped hence only particular states are suitable for storage.  For example, in $^{87} \text{Rb} $, the typical states that are used are the $ F= 1, m_F=-1 $ and $ F=2, m_F=1 $ states as they are magnetically trapped and have the same response to magnetic field fluctuations which reduce dephasing effects \cite{bohi2009}, \cite{riedel2010}, \cite{byrnes2014}. In such systems the coherent control is performed using microwave and radio frequency control.  This can be combined with non-destructive measurement methods such as that developed in Refs. \cite{ilookeke14} and \cite{ilookeke2016} to perform the detection.

The interaction between the BEC and the light is described by
\begin{eqnarray*}
\hat{H}_{\mathrm{I}}  &=&  -\sum_{j=1}^2 \int d\emph{\textbf{x}} \Bigg\{ \hat{\psi}_{g j}^{\dag}(\emph{\textbf{x}}) \left[ \emph{\textbf{d}}_{j}\cdot  \hat{\emph{\textbf{E}}}(\emph{\textbf{x}}) \right] \hat{\psi}_{e j}(\emph{\textbf{x}})+\\
&&\hat{\psi}_{e j}^{\dag}(\emph{\textbf{x}}) \left[ \emph{\textbf{d}}_{j}\cdot \hat{\emph{\textbf{E}}}(\emph{\textbf{x}}) \right] \hat{\psi}_{g j}(\emph{\textbf{x}}) \Bigg\},
\end{eqnarray*}
where $\emph{\textbf{d}}_{j}$ is the transition dipole moment of atoms in the $j$th component, and the electric field operator is
\begin{eqnarray}
\hat{\emph{\textbf{E}}}(\emph{\textbf{x}}, t)  &=&  i \sum_{\bm{k},\sigma} \sqrt{\frac{\hbar \omega_k}{2\epsilon_0 V }} \left( \hat{a}_{\bm{k} \sigma}e^{i \bm{k} \cdot \emph{\textbf{x}}-i\omega_kt} - \mathrm{H.c.}\right),
\label{efieldop}
\end{eqnarray}
where $V$ is the volume of quantization, $\omega_k$ is the frequency of the light of wavenumber $ k = |\bm{k} | $,  $\sigma$ labels the polarization of the light, and $ \hat{a}_{\bm{k} \sigma} $ is a photon annihilation operator for wavenumber $ k $ and polarization $ \sigma $.  The electromagnetic degrees of freedom have the Hamiltonian in this basis
\begin{align}
\hat{H}_{\mathrm{E}}  =  \sum_{\bm{k},\sigma} \hbar \omega_k \left[  \hat{a}_{\bm{k}  \sigma}^\dagger  \hat{a}_{\bm{k}  \sigma} + \frac{1}{2} \right]  .
\end{align}

The total Hamiltonian of the whole system is
\begin{eqnarray}
\hat{H}_{all}=\hat{H}_{\mathrm{S}}+\hat{H}_{\mathrm{E}}+\hat{H}_{\mathrm{I}} .
\end{eqnarray}
We may then use adiabatic elimination to eliminate the excited state $ e $ to obtain an effective Hamiltonian that only involves the ground state. Moving to the interaction picture, and making a rotating-wave approximation to remove terms that do not conserve energy, we obtain the effective interaction Hamiltonian
\begin{eqnarray}
\label{new-H}
\hat{H}_{\mathrm{eff}}=-2 \sum_{j=1}^2 \int d \emph{\textbf{x}} \hat{\psi}_{g j}^{\dag}(\emph{\textbf{x}})
\frac{| \bm{d}_j \cdot \hat{\bm{E}}^- |^2 }{\Delta_j}
 \hat{\psi}_{g j}(\emph{\textbf{x}}) ,
\end{eqnarray}
where  $\hat{\bm{E}}^-= \hat{a} E^-(\emph{\textbf{x}}) \varepsilon $ are the negative frequency components of (\ref{efieldop}) ($\varepsilon$ is the photon polarization) and the detuning for the $ j$th level is
\begin{align}
\Delta_j=\omega_{e j}-\omega_{g j}-\omega,
\end{align}
 where $\omega$ is  the frequency of the incident light and we have made a single-mode approximation.

%
The dynamics of a two-component spinor BEC interacting with external fields can be described  a unitary evolution characterized by a unitary operator (see Sec. 11.2.3 of Ref. \cite{GZ04} and Ref. \cite{BHJ2007})
\begin{eqnarray}\label{Uni1}
\!dU_t \!=&&\Bigg\{\!\sum_{j=1}^2\int\!\! dx  \hat{L}_j(x) dB^{\dag}_{j}(t)\! - \!\sum_{j=1}^2\int\! dx  \hat{L}_j^{\dag}(x) dB_{j}(t) \nonumber\\
 &&-\sum_{j=1}^2\int dx\frac{1}{2} \hat{L}_j^{\dag}(x)  \hat{L}_j(x)dt +i\hat{H} dt\Bigg\} U_t,
\end{eqnarray} with the measurement operator given by
\begin{align}\label{Uni}
 \hat{L}_{j}(x)=\int dy\hat{\psi}_{g j}^{\dag}(y)\hat{\psi}_{g j}(y)\chi(x-y)_j,
\end{align}
and $B_{j}$ are boson fields in the environment, and $\chi(x)_j$ is a kernel function and its width is related to the resolution length scale  of the measurement of the local particle density operator $\hat{\psi}_{g j}^{\dag}(x)\hat{\psi}_{g j}(x)$.

Under continuous PCI detection, we can continuously monitor the observable
\begin{eqnarray}
\label{PCI-measurement}
Y_j(t) = B_j(t) + B_j^{\dag}(t),
\end{eqnarray} where the measurement $Y(t)$ satisfies $[Y_j(t), Y_j(s)]=0$ for all $s, t\geq0$.
In the Heisenberg picture, $h_t(X) = U^{\dag}_t X U_t$ denotes the evolution of a system observable $X$  for any operator and satisfies $[h_t(X), Y_j(s)]=0$ for all $s, t\geq0$, see \cite{BVP94} for details.

 Then, from \eqref{PCI-measurement},  any  observable $X$ of a two-component spinor BEC considered in this paper can be best estimated by \cite{BHJ2007}
\begin{eqnarray}
\label{quantum-estimation}
      \!\!d\pi_t(X)\!=&&\pi_t\left(\mathcal{L}(X)\right)dt\!+\!\!\sum_{j=1}^2\!\int\! d\emph{\textbf{x}}\Bigg\{\!\pi_t\!\left(\! \hat{L}_j^{\dag}(\emph{\textbf{x}})X\!+\!X \hat{L}_j(\emph{\textbf{x}})\right)\!-\nonumber\\
&&\pi_t\!\left(\! \hat{L}_j^{\dag}(\emph{\textbf{x}})+ \hat{L}_j(\emph{\textbf{x}})\right)\pi_t(X)\Bigg\}dW_j(\emph{\textbf{x}},t),
\end{eqnarray} where $\pi_t(X)$ is the conditional expectation of $X$  and the Lindblad operator $\mathcal{L}(X)$ is given by
\begin{eqnarray}\label{Lindblad}
\mathcal{L}(X)=&&i[\hat{H}, X]+\sum_{j=1}^2\!\int\!\!\! d\emph{\textbf{x}}\Bigg\{\! \hat{L}_j^{\dag}(\emph{\textbf{x}})X \hat{L}_j(\emph{\textbf{x}})-\nonumber\\
&&\frac{1}{2}\left( \hat{L}_j^{\dag} \hat{L}_jX+X \hat{L}_j^{\dag} \hat{L}_j\right)\Bigg\}
\end{eqnarray} and classical Wiener
increment $dW_j(\emph{\textbf{x}},t)=dY_j(t)-\pi_t\!\left(\! \hat{L}_j^{\dag}(\emph{\textbf{x}})+ \hat{L}_j(\emph{\textbf{x}})\right)dt$.

Let $\rho_c$ be the conditional density operator satisfying $\pi_t(X)=\mathrm{Tr}(X\rho_c)$.  Then, based on a system-bath interaction between the BEC and the electric field \cite{szigeti2009},  we have the
following conditional master equation with external feedback control Hamiltonian $H_f$:
\begin{eqnarray}
\label{conditional-master-1}
d\rho_c&=&-i[\hat{H},\rho_c]dt+\sum_{j=1}^2\alpha_{j}\int d \emph{\textbf{x}}\mathcal{D}[\mathcal{\hat{L}}_{j}(\emph{\textbf{x}})]\rho_cdt+ \nonumber\\
&&\sqrt{\eta}\sum_{j=1}^2\sqrt{\alpha_{j}}\int d \emph{\textbf{x}}\mathcal{H}[\mathcal{\hat{L}}_{j} (\emph{\textbf{x}})]  \rho_c  dW_j(\emph{\textbf{x}},t),
\end{eqnarray}
where $\hat{H}=\hat{H}_{\mathrm{eff}}+H_f$, $\mathcal{D}[c]\rho_c = c\rho_cc^{\dag}-\frac{1}{2}(c^{\dag}c\rho_c + \rho_cc^{\dag}c)$ and $\mathcal{H}[c]\rho_c=c\rho_c+\rho_cc^{\dag}-\mathrm{Tr}((c+c^{\dag}) \rho_c)\rho_c $. $\eta$   is the detection efficiency and $\alpha_{j}$ is the
effective interaction strength corresponding to $W_j$. $\mathcal{\hat{L}}_{j}$ is the measurement operator of  the  analogous form \eqref{Uni}.

\subsection{Total spin approximation}

Eq. (\ref{conditional-master-1}) describes the dynamics of the BEC for an arbitrary spatial wavefunction. For low temperatures, small excited state populations, and for atomic species such as $^{87}\mbox{Rb}$ where the inter and intra atomic scattering lengths are approximately equal, the spatial wavefunction can be taken to be the same for all the components.  This will allow us to eliminate the spatial degrees of freedom giving a master equation just for the spin.

Let us now change the basis of the boson operators in terms of the eigenstates of (\ref{singleparticleham}) and then we have
\begin{align}
\hat{\psi}_{n j} (\bm{x}) = \sum_l c_{n j l}  \psi_{n j l}(\bm{x}),
\end{align}
where $c_{n j l} $ is a bosonic annihilation operator for a state in level $ n = g,e $, hyperfine state $ j $, and $ l$th eigenstate of (\ref{singleparticleham}). Assuming that all the atoms occupy the ground state $ l = 0 $ in this expansion, and taking $ \psi_{g j 0}(\bm{x}) = \psi_0(\bm{x}) $ and $ b_j = c_{g j 0} $, we have
\begin{eqnarray}
\label{spin-form}
\hat{\psi}_{g j}(\bm{x})=b_j\psi_0(\bm{x}).
\end{eqnarray}
The $ b_j $ satisfy the usual bosonic commutation relations $\left[b_j, b_k^\dagger\right ] =\delta_{jk} $.
Substituting \eqref{spin-form} into \eqref{new-H}, we have the effective interaction Hamiltonian in
terms of the relative population difference between two components given by
\begin{eqnarray}
\hat{H}_{\mathrm{eff}}   &=& G_1 b_{1}^{\dag}b_{1}+ G_2 b_{2}^{\dag}b_{2},
\end{eqnarray}
where the coefficients  for $ j = 1,2 $ are
\begin{align}
G_j = - \frac{2}{\Delta_j} \int d\emph{\textbf{x}} | E^- ( \emph{\textbf{x}}) \psi_0(\emph{\textbf{x}}) \langle g j | \bm{d}_j \cdot \varepsilon | e j \rangle |^2
\end{align}
We now define spin operators $ S^{x,y,z} $ and particle number operator $ N $ as
\begin{align}
S^x & =  b_{1}^{\dag}b_{2} + b_{2}^{\dag}b_{1},\nonumber \\
S^y & =   -i b_{1}^{\dag}b_{2} + i b_{2}^{\dag}b_{1}, \nonumber \\
S^z & =  b_{1}^{\dag}b_{1} - b_{2}^{\dag}b_{2}, \nonumber \\
N & =  b_{1}^{\dag}b_{1} + b_{2}^{\dag}b_{2}.
\end{align}
These operators satisfy the commutation relations
\begin{align}
\label{relation-bec}
[S^j,S^k] = 2i\epsilon_{jkl}S^l,
\end{align}
where $\epsilon_{jkl}$ is the
Levi-Civita antisymmetric tensor.  The effective Hamiltonian can then be written
\begin{align}
\hat{H}_{\mathrm{eff}}  = G S^{z}+g N,
\label{spin-form-H}
\end{align}
where the coefficients
\begin{align}
 G & =\frac{G_1-G_2}{2}, \nonumber \\
 g & = \frac{G_1+G_2}{2} .
\end{align}
Under this approximation the measurement operator is
\begin{eqnarray*}
\mathcal{\hat{L}}_{j}(\emph{\textbf{x}})=M_j(\emph{\textbf{x}}) n_j,
\label{measop}
\end{eqnarray*}
where the number operator  $ n_j = b_{j}^{\dag}b_{j} $ and the coefficient $M_j$ is given by
\begin{align}
M_j(\emph{\textbf{x}})=\int d\emph{\textbf{x}}' |\psi_0(\emph{\textbf{x}}') |^2 \chi_j(\emph{\textbf{x}}-\emph{\textbf{x}}')
\end{align}
is a real function. The superoperator then is
\begin{align}
\mathcal{D}[\mathcal{\hat{L}}_{j}(\emph{\textbf{x}})]\rho_c = &
\frac{M_j^2 (\emph{\textbf{x}}) }{2} \left[ 2  n_j \rho_c n_j
-  n_j^2 \rho_c - \rho_c n_j^2  \right] \\
 =&\frac{1}{4} M_j^{2}(\emph{\textbf{x}})\mathcal{D}[S^{z}]\rho_c.
\label{dexpr}
\end{align}
where we have assumed that the total number of atoms $N$ is a constant. Similarly,
\begin{align}
\mathcal{H}[\mathcal{\hat{L}}_{j}(\emph{\textbf{x}})]\rho_c  = & M_j(\emph{\textbf{x}}) \Big[ n_j \rho_c  +
\rho_c n_j -  2 \mathrm{Tr}(n_j \rho_c)\rho_c \Big]\\
= & -\frac{(-1)^j}{2} M_j(\emph{\textbf{x}})\mathcal{H}[S^{z}]\rho_c.
\label{hexpr}
\end{align}
Substituting (\ref{dexpr}) and (\ref{hexpr}) into the conditional master equation
(\ref{conditional-master-1}) we obtain
\begin{align}
\label{conditional-master1}
d\rho_c=& -i[GS^{z}+gN+H_f,\rho_c]dt+A\mathcal{D}[S^{z}]\rho_cdt\nonumber \\
& +\sqrt{\eta} B\mathcal{H}[S^{z}]\rho_c dw(t),
\end{align}
where the coefficients $A$ and $B$ are given by
\begin{align}
A= & \sum_{j=1}^2 \frac{\alpha_{j}}{4}\int  M_j^2 (\emph{\textbf{x}}) d\emph{\textbf{x}}, \\
B= & \sqrt{A} ,
\end{align}
and the noise operators operating on the spins are normalized as
\begin{align}
dw = \frac{\sqrt{\alpha_{1}}  d\tilde{W}_{1}- \sqrt{\alpha_{2}} d\tilde{W}_{2}}{2B},
\end{align}
where $d\tilde{W}_j(t)= \int d\emph{\textbf{x}} M_j(\emph{\textbf{x}})dW_j(\emph{\textbf{x}},t)$.

%

Eq. (\ref{conditional-master1}) completes our derivation of the conditional master equation for a BEC under continuous measurement.  In the above, while we specialized our derivation for BECs, the identical equation holds true for ensembles.  A similar argument can be performed for ensembles, instead of BECs, which is presented in the Appendix.  In order for this approximation to hold, we require that any operations on the ensemble are symmetric under particle interchange. Specifically, the PCI measurement and control Hamiltonian should be symmetric under interchange of atoms in the ensemble.  Due to the relatively small size of typical ensembles compared to typical laser pulses, this is a reasonable approximation as long as the column density of the ensemble is small enough such that the couplings between all the atoms are the same. We note that it is important to consider the many particle nature of the ensemble as we deal with collective operations and measurements throughout the feedback process.  A collective measurement on an ensemble behaves differently to that of a single qubit.  The optimum fidelity one can estimate an unknown quantum state approaches 1 as $ \propto 1/N $ \cite{massar1995}, and similarly for nondestructive measurements \cite{ilookeke14}.  This means that for our feedback control it should be possible to obtain a better estimate for the state as $ N $ increases, and thereby improving the feedback control.

\section{Quantum control methods}
\label{sec:num}
Quantum control strategies consists of two steps: an estimation step and a control step \cite{WM2009}, \cite{BHJ2007}. In the estimation step, the state estimate which
results from equation \eqref{quantum-estimation} can then be used to form the  control Hamiltonian  that can modify the system
Hamiltonian in order to achieve the desired control of the quantum system. In this section, we come to the control step and aim to find control strategies to  manipulate quantum systems in real time.

%

The design of  the control Hamiltonian is based on the invariant set theorem (LaSalle's invariance principle) that is a criterion for the asymptotic stability of a  nonlinear dynamical system and provides  a useful tool to analyze convergence to a desired state \cite{LL1961}.
Let us  investigate the dynamics of the equation \eqref{conditional-master1}, first without feedback ($H_f=0$). Consider the quantity
\begin{align}
{\cal S}(\rho_c)=\mathrm{Tr}[(S^z)^2\rho]-(\mathrm{Tr}[(S^z)\rho])^2
\end{align}
in \eqref{conditional-master1}. It is easily obtained that
\begin{eqnarray}\label{result}
\mathcal{A}{\cal S} (\rho_c)=-4B^2\eta {\cal S}(\rho)^2\leq0,
\end{eqnarray}where $\mathcal{A}$ is infinitesimal generator of $\rho_c$.
Note that $\mathbb{E}[{\cal S}(\rho^2_t)]\geq0$, hence from \eqref{result} we conclude that $\mathbb{E}[{\cal S}(\rho_c)]$ decreases monotonically. 
Therefore, ${\cal S}(\rho_c)$ converges to $0$ as $t$ goes to $\infty$. But the only states $\rho$ satisfying ${\cal S}(\rho)=0$ are the eigenstates of $S^z$, which implies that the state $\rho$ governed by \eqref{result} with $H_f=0$ must collapse onto one of the eigenstates of $S^z$. From a physical point of view, this is the expected result as the PCI performs a measurement in the $S^z$ basis, and hence an arbitrary initial state is driven towards $S^z$ eigenstates.

Let us now consider the case that $H_f\neq 0$.  What we would like to achieve is the solution to the following problem: for the time evolution of a state given by \eqref{conditional-master1}, find the control Hamiltonian $H_f$ that drive an unknown state  into a desired state $\rho^f$. To this end, define a non-negative continuous function that represents the distance between a state $\rho$ and a desired state $\rho_f$ given by
\begin{eqnarray}
\label{lypunovfun}
V(\rho, \rho_f)=\frac{1}{2}\parallel \rho-\rho_f\parallel ^2=\frac{1}{2}\mathrm{Tr}[(\rho-\rho_f)^2] .
\end{eqnarray}
If we can find control Hamiltonian of the form
\begin{eqnarray}
\label{controlham}
H_f=\sum_{k=x, y, z}u_k H_k,
\end{eqnarray}
where $u_k$ is external control signal and $H_k=S^k$ is a time-independent Hamiltonian operator that guarantees that
\begin{align}
\mathcal{A}V(\rho, \rho_f)\leq 0,
\end{align}
this solves our problem.


\section{Numerical simulation}
\label{Simulation}

In this section we test the control Hamiltonian (\ref{controlham}) in the conditional master equation (\ref{conditional-master1}) to evaluate the performance of preparing various quantum state. We will assume that we have a feedback control Hamiltonian of the form \eqref{controlham} that performs global spin rotations on the ensemble or BEC.
Starting from  (\ref{conditional-master1}) we derive the following nonlinear stochastic equations by multiplying by total spin operators $ S^{x,y,z} $ and taking the trace,
\begin{align}
\langle X \rangle_t= \mathrm{Tr} ( \rho_c X ) ,
\end{align}
where $ X $ is an arbitrary operator.  Defining normalized operators and control signals
\begin{align}
s^{x,y,z} \equiv \frac{S^{x,y,z}}{N},
\end{align}
we obtain evolution equations linear in these variables as
\begin{align}
d\langle s^x \rangle_t= &2 \Big(\!u_y\langle s^z \rangle_t\!-\!u_z\langle s^y \rangle_t\!-\! A\langle s^x \rangle_t -G\langle s^y \rangle_t\Big) dt \nonumber\\
&+ B N \sqrt{\eta} \Big( \langle s^x s^z \rangle_t+\langle s^z s^x \rangle_t-2 \langle s^z \rangle_t\langle s^x \rangle_t \Big) dw,
\nonumber \\
d\langle s^y \rangle_t= & 2 \Big( u_z\langle s^x \rangle_t\!-\! A\langle s^y \rangle_t\!-\! u_x\langle s^z \rangle_t + G\langle s^x \rangle_t \Big)  dt  \nonumber\\
&+ BN \sqrt{\eta} \Big( \langle s^y s^z \rangle_t + \langle s^z s^y \rangle_t - 2\langle s^z \rangle_t\langle s^y \rangle_t \Big) dw,
\nonumber \\
 d\langle s^z \rangle_t  = &  2 \Big( \!u_x\langle s^y \rangle_t\!- u_y\langle s^x \rangle_t \Big) dt+\nonumber\\
 &2 BN \sqrt{\eta}(\langle {(s^z)}^2  \rangle_t- \langle s^z \rangle_t^2) dw,
\label{singlespinevolution}
\end{align}
These equations involve higher powers of total spin operators, which themselves have time evolution equations
\begin{align}\label{twospineqns}
d\langle s^x s^z \rangle_t =&2 \Big( u_x\langle s^x s^y \rangle_t+u_y\langle {s^z}^2  \rangle_t-u_y \langle {s^x}^2  \rangle_t  - u_z\langle s^y s^z \rangle_t \nonumber\\
& - A\langle s^x s^z \rangle_t - G\langle s^y s^z \rangle_t \Big) dt,\nonumber\\
d\langle s^y s^z \rangle_t =&2 \Big( u_x\langle {s^y}^2  \rangle_t-u_x\langle {s^z}^2  \rangle_t-u_y\langle s^y s^x  \rangle_t+u_z\langle s^x s^z \rangle_t  \nonumber\\
&  - A\langle s^y s^z \rangle_t + G\langle s^x s^z \rangle_t\Big) dt,
\nonumber\\
d\langle s^x s^y  \rangle_t =&2 \Big( -u_x\langle s^x s^z \rangle_t + u_y\langle s^z s^y \rangle_t + u_z\langle {s^x}^2  \rangle_t  \nonumber\\
&- u_z\langle {s^y}^2  \rangle_t  -2A\langle s^y s^x  \rangle_t -2A \langle s^x s^y  \rangle_t \nonumber \\
&  + G \langle {s^x}^2  \rangle_t - G\langle {s^y}^2  \rangle_t \Big)dt
\end{align}
where the equations for the complex conjugates can be found from $\langle s^i s^j \rangle= \langle s^j s^i\rangle^* $ for $i, j=x, y, z$. We also have
\begin{align}
d\langle {s^x}^2  \rangle_t&  =  2  \Big(2A\langle {s^y}^2  \rangle_t -2A \langle {s^x}^2  \rangle_t  + u_y\langle s^x s^z \rangle_t  + u_y\langle s^z s^x \rangle_t \nonumber\\
& - u_z\langle s^x s^y  \rangle_t  -u_z\langle s^y s^x \rangle_t - G \langle s^x s^y  \rangle_t - G\langle s^y s^x \rangle_t\Big) dt \nonumber\\
d\langle {s^y}^2  \rangle_t& = 2 \Big( 2A\langle {s^x}^2  \rangle_t- 2A \langle {s^y}^2 \rangle_t  -u_x\langle s^y s^z \rangle_t - u_x\langle s^z s^y \rangle_t    \nonumber\\
&  + u_z\langle s^x s^y  \rangle_t  + u_z\langle s^y s^x \rangle_t + G\langle s^x s^y  \rangle_t+ G \langle s^y s^x \rangle_t \Big) dt \nonumber \\
d\langle {s^z}^2 \rangle_t&  =  2  \Big(u_x\langle s^y s^z \rangle_t + u_x\langle s^z s^y \rangle_t - u_y\langle s^x s^z \rangle_t \nonumber\\
& -  u_y\langle s^z s^x \rangle_t \Big) dt .
\end{align}
In (\ref{twospineqns}) there are only second order correlations in the spin operators
as we have made the approximation
\begin{align}
\langle s^i s^j s^k \rangle \approx \langle s^i s^j \rangle \langle s^k \rangle \approx \langle s^i  \rangle \langle s^j s^k \rangle,
\end{align}
which neglects third order correlations between spin operators.  As we are primarily interested in evolving the system towards a desired $ (\langle s^x \rangle, \langle s^y \rangle, \langle s^z \rangle) $ position on the Bloch sphere, third order correlations should play a negligible role in determining the position.  This also has the effect of neglecting the stochastic terms on the second order evolution equations.

\begin{figure}[t]
\includegraphics[width=\columnwidth]{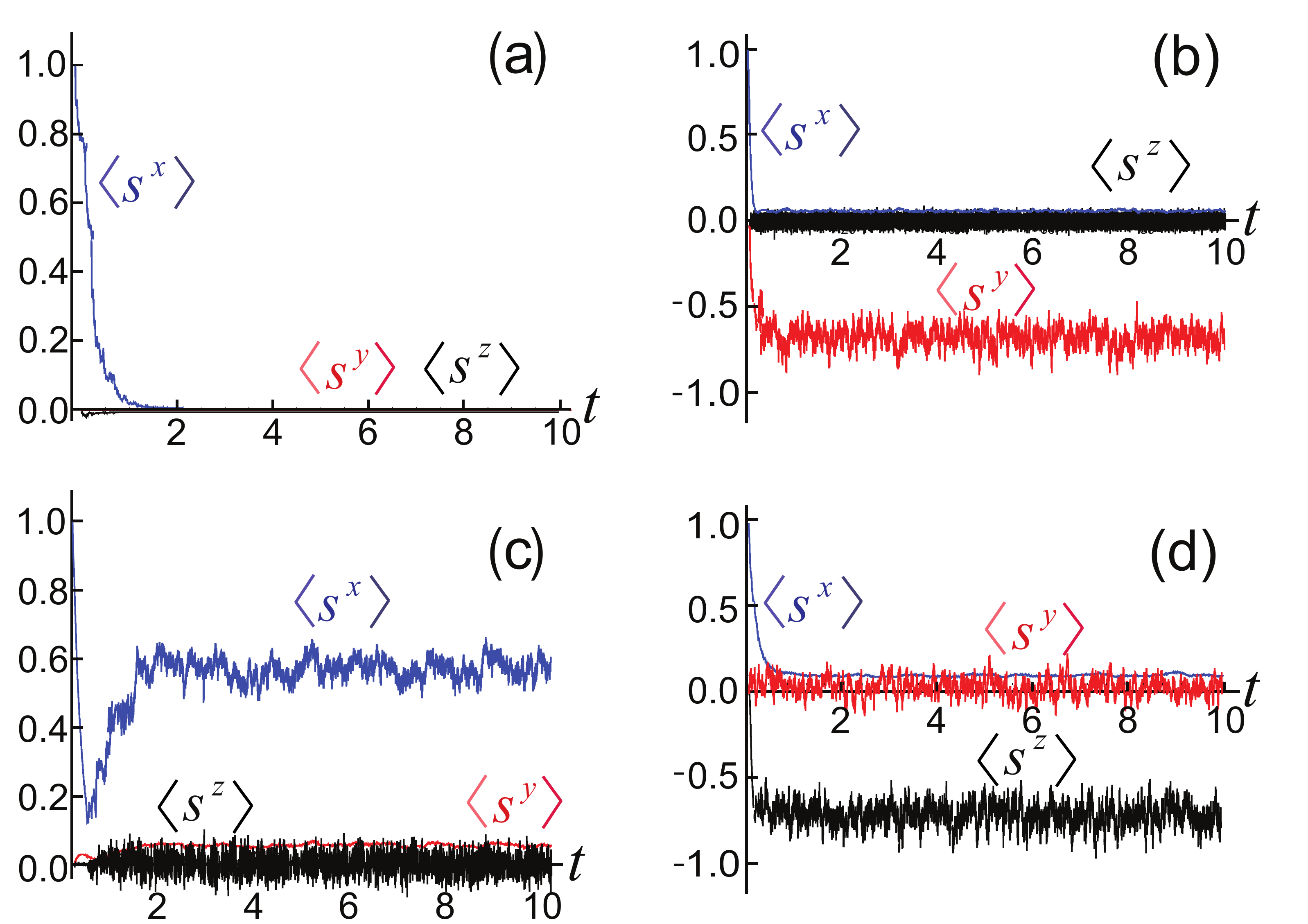}
\caption{(Color online) Typical trajectory of $\langle s^{x,y,z} \rangle_t $ under evolution equations (\ref{singlespinevolution}) with $N=100$ and $\eta=1$. Common parameters are  $G=0.0001$, $B^2=A = 0.04$, $\Delta t =1$. (a) Without feedback control $ H_f = 0$;  (b) with feedback $u_x = 9.5 \langle s^z \rangle_t $, $u_y=0$, $u_z=0$; (c) $u_x = 0$, $u_y = 0.01+8\langle s^z \rangle_c $, $u_z =0$; (d) $u_x = 0$, $u_y=0$, $u_z = 6\langle s^y \rangle_t $. \label{fig2}}\vspace{-1em}
\end{figure}

For the new evolution equations \eqref{singlespinevolution}, we need to redefine the function \eqref{lypunovfun} in terms of $\langle s^{(x,y,z)} \rangle$ given by
\begin{eqnarray}
\label{new-lypunovfun}
V(\bm{s}_t, \bm{s}_f)=\frac{1}{2}(\bm{s}_t-\bm{s}_f) \cdot (\bm{s}_t-\bm{s}_f)=\frac{1}{2}\parallel \bm{s}_t-\bm{s}_f \parallel^2, \nonumber \\
\end{eqnarray}
where the  column vector $\bm{s}_t=[\langle s^x \rangle_t,  \langle s^y \rangle_t,  \langle s^z \rangle_t]^T$  specifies a point on the unit sphere  and $\bm{s}_f=[\langle s^x \rangle_f ,  \langle s^y \rangle_f ,  \langle s^z \rangle_f]^T$ represents a desired point.

To ensure $\mathcal{A}V(\bm{s}_t, \bm{s}_f)\leq 0$, we choose our control laws as
\begin{align}
u_x & = \xi_x+ \beta_{xx} \langle s^x \rangle_t + \beta_{xy}  \langle s^y \rangle_t + \beta_{xz} \langle s^z \rangle_t,  \nonumber \\
u_y & = \xi_y+ \beta_{yx}\langle s^x \rangle_t + \beta_{yy} \langle s^y \rangle_t + \beta_{yz}\langle s^z \rangle_t,  \nonumber \\
u_z & = \xi_z+ \beta_{zx}\langle s^x \rangle_t +\beta_{zy} \langle s^y \rangle_t +\beta_{zz}\langle s^z \rangle_t,
\end{align}
where the $ \xi_k $ and $ \beta_{kl} $ are constants throughout the feedback evolution.

The initial state is chosen as a spin coherent state polarized in the $ S^x $ direction \cite{byrnes2012}
\begin{align}
| t = 0 \rangle = \frac{1}{\sqrt{N!}}\left(\frac{b_1^{\dag}+ b_2^{\dag}}{\sqrt{2}} \right)^{N}|0\rangle,
\end{align}
where we assume that the particle number $ N $ is a constant henceforth.  In terms of the normalized spin variables, the initial states $ t = 0 $ are therefore
\begin{align}
\langle s^x \rangle_0  & =\langle {s^x}^2 \rangle_0=1, \nonumber \\
\langle s^y \rangle_0  & = \langle s^z \rangle_0  = \langle s^x s^z \rangle_0  = \langle s^x s^y \rangle_0 = 0, \nonumber \\
\langle s^y s^z \rangle_0  & = i , \nonumber \\
\langle {s^y}^2 \rangle_0 & = \langle {s^z}^2 \rangle_0 = \frac{1}{N} .
\label{initconds}
 \end{align}
The equations are evolved using the stochastic differential equation solver in Mathematica with time step $ \Delta t/\tau_0 $.  Here, the timescale of the evolution is taken to be $ \tau_0 = 1/G $.

\begin{figure}[t]
\includegraphics[width=\columnwidth]{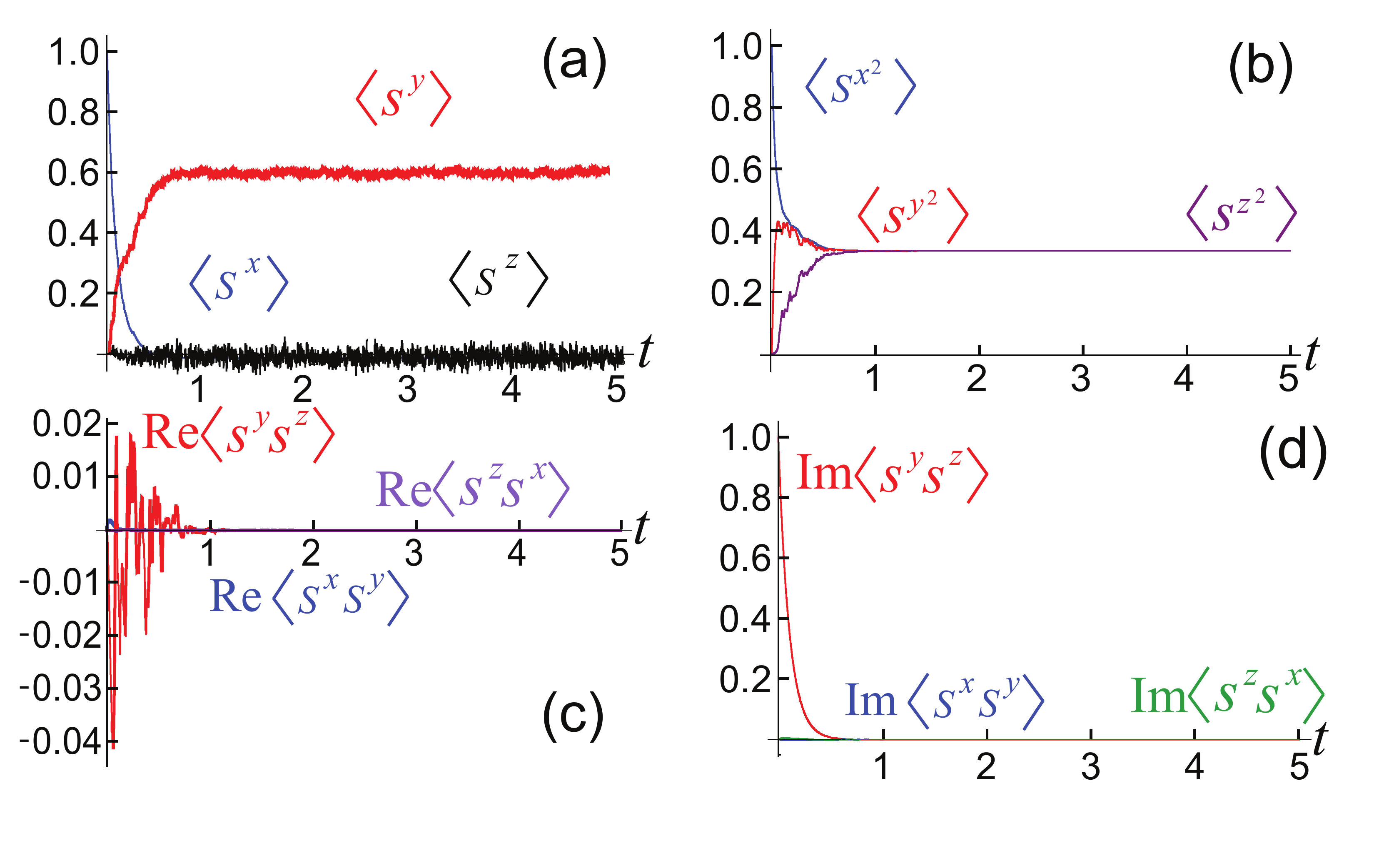}
\caption{Trajectory correlations under evolution equations (\ref{singlespinevolution}) with  $N=100$ and $\eta=1$. The control law is chosen as $u_x = -14.5\langle s^z \rangle_t $, $u_y =0$, $u_z =0$ and  parameters $G=0.0001$, $B^2=A = 0.04$, $\Delta t =1$. Correlations shown are for (a)  $\langle s^{x,y,z} \rangle_t $; (b) $\langle {s^{x,y,z}}^2 \rangle_t $; (c) real parts of $\langle s^{j}s^k  \rangle_t  $; (d)  imaginary parts $\langle s^{j} s^k \rangle_t $.  \label{fig3}}
\end{figure}

In Fig. \ref{fig2}(a) we plot typical trajectories of the the normalized spin expectations $\langle s^{x,y,z} \rangle $ without any control law.  Starting from the initial condition (\ref{initconds}), We see that all the spin expectations quickly decay to zero, indicating that no polarized state can be maintained with the continuous measurement in place.  At the level of the equations it is clear that this occurs due to the term proportional to $ A $ in (\ref{singlespinevolution}), which causes exponential decay of the initial $\langle s^{x} \rangle $  polarization.  This term is a Lindblad dephasing term in the master equation (\ref{conditional-master1}) which destroys any coherence and collapses a state onto the $ S^z $ basis.


Including the feedback control, it is possible to stabilize a non-zero value of  $\langle s^{x,y,z} \rangle_t $ at steady state.  Figure \ref{fig3}  shows an example of the stabilization of $\langle s^{y} \rangle_t $ as well as the other correlations that are involved in the time evolution.
It can be seen from Fig. \ref{fig3} that  $\langle s^j \rangle_t $, $\langle {s^j}^2 \rangle_t $ and $\langle s^j s^k \rangle_t  $ ($j,k=x,y,z $) arrive at their final values at the same time. We see that the cross correlations $\langle s^j s^k \rangle_t  $ all are zero at steady state, which is primarily due to the fact that $ \langle s^{x,z} \rangle \approx 0 $ for this case, which makes these correlations zero.  The fact that $ \langle {s^{x}}^2 \rangle, \langle {s^{z}}^2 \rangle $ are non-zero at steady state is consistent with spin coherent states polarized in the $ s^{y} $ direction, which have quantum noise in the $ s^{x,z} $ directions.

By using other control laws it is possible to stabilize the spin at other locations on the Bloch sphere.  First let us consider the effect of control acting only on the $ S^x $ term in $ H_f $, specifically  $u_x = \xi_x+ \beta_{xz} \langle s^z \rangle $ and $u_y=u_z= 0$.  In Fig. \ref{fig2}(b) we see that such control laws  only influence $ \langle s^y \rangle$ while $\langle s^x \rangle $ and $\langle s^z \rangle$ fluctuate around $0$. By using various combinations of feedback parameters, it is possible to control all the $ \langle s^{x,y,z} \rangle $ components.  Using a control law such as $u_y = \xi_y +\beta_{yz} \langle s^z \rangle $ allows control of $ \langle s^z \rangle_c $ as shown in Fig. \ref{fig2}(c).  A combination of feedback on all $ S^{x,y,z} $ in $ H_f $ allow for the control of multiple components simultaneously, as shown in Fig. \ref{fig2}(d).

\begin{figure}[t]
\includegraphics[width=\columnwidth]{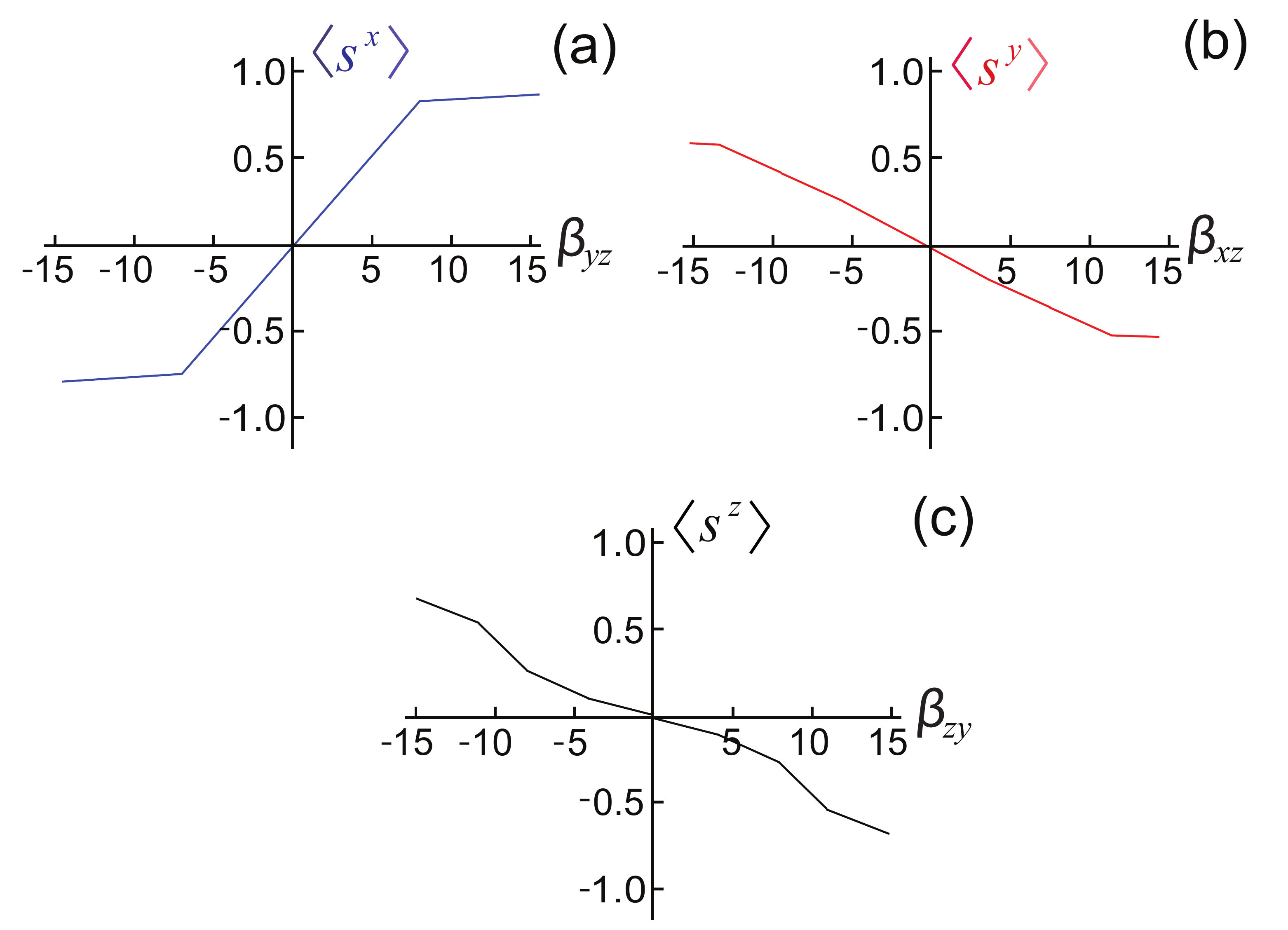}
\caption{Steady state values of $ \langle s^{x,y,z} \rangle $ as a function of feedback control parameters $ \beta_{kl} $. (a) $u_y = \xi_y+\beta_{yz} \langle s^z \rangle_t $ and $u_x= u_z=0$; (b) $u_x = \xi_x+\beta_{xz} \langle s^z \rangle_t $ and $u_y= u_z=0$; (c) $u_z = \xi_x+\beta_{zy} \langle s^y \rangle_t $ and $u_x= u_y=0$. \label{fig4}}
\end{figure}

We may obtain the steady state values by averaging over a time $ t = 10$ after the initial transient dynamics is finished.  The averaging process reduces the noise fluctuations.
The steady state values for various types of feedback control (different feedback parameters) are shown in Fig. \ref{fig4}. We see that there are several combinations which are particularly effective at driving the spin in certain directions. It is easily checked that the above combinations ensures that $\mathcal{A}V(\bm{s}_t, \bm{s}_f)\leq 0$ holds.  For instance, in Fig. \ref{fig4}(b) we see that feeding back the estimate of $s^z $ signal along the coordinate $s^x $ is effective to drive the $\langle s^y \rangle $  component to a desired value by applying different values of $\beta_{xz}$. A brief summary of the feedback signal, control law, and the desired position value on the Bloch sphere is shown in Table \ref{tab1}. Combining such control laws can give the stabilization of an arbitrary point on the Bloch sphere.

\begin{table}[b]
\!\!\!\begin{tabular}{c c|c c c c|c c|}
\cline{3-8}
\multicolumn{2}{ c| }{} & \multicolumn{4}{ c| }{Feedback signal}&\multicolumn{2}{c|}{Desired $(\langle s^x \rangle,\langle s^y \rangle, \langle s^z \rangle) $ } \\
\multicolumn{2}{ c| }{} & $1 $  & $\langle s^x \rangle_t $ & $\langle s^y \rangle_t $ & $\langle s^z \rangle_t$& \multicolumn{2}{c|}{the
Bloch sphere position}  \\
\cline{1-8}
\multicolumn{1}{ |c }{\multirow{4}{*}{Control law} } & \multicolumn{1}{ c| }{$ u_x $} & 0 & $0$ & $0$ & $ \pm $& \multicolumn{2}{ c| } { } \\
\multicolumn{1}{ |c }{} & \multicolumn{1}{ c| }{$ u_y $} & 0 & $0$ & $0$ & $ 0 $ & \multicolumn{2}{ c| } {$\left(0, \mp, 0\right)$ } \\
\multicolumn{1}{ |c }{} & \multicolumn{1}{ c| }{$ u_z $} & 0 & $0$ &  $ 0 $ & 0& \multicolumn{2}{ c| } {} \\
\cline{1-8}
\multicolumn{1}{ |c }{\multirow{4}{*}{Control law} } & \multicolumn{1}{ c| }{$ u_x $} &0 & $0$ & $0$ & $ 0 $& \multicolumn{2}{ c| } {  } \\
\multicolumn{1}{ |c }{} & \multicolumn{1}{ c| }{$ u_y $} & 0 & $0$ & $0$ & $ \pm $ & \multicolumn{2}{ c| } { $\left(\pm, 0, 0\right)$  } \\
\multicolumn{1}{ |c }{} & \multicolumn{1}{ c| }{$ u_z $} & 0 & $0$ &  $ 0 $ & 0& \multicolumn{2}{ c| } { } \\
\cline{1-8}
\multicolumn{1}{ |c }{\multirow{4}{*}{Control law} } & \multicolumn{1}{ c| }{$ u_x $} &0 & $0$ & $0$ & $0 $& \multicolumn{2}{ c| } { } \\
\multicolumn{1}{ |c }{} & \multicolumn{1}{ c| }{$ u_y $} & 0 &$0$ & $0$ & $ 0 $ & \multicolumn{2}{ c| } {$(0, 0, \mp)$  } \\
\multicolumn{1}{ |c }{} & \multicolumn{1}{ c| }{$ u_z $} & 0 &$0$ &  $ \pm $ & 0& \multicolumn{2}{ c| } {  } \\
\cline{1-8}
\end{tabular}
\caption{A summary of the effect of various combinations of feedback signal (columns), the feedback control law along suitable coordinates (rows). \label{tab1}}
\end{table}

\section{Conclusions}
\label{sec:conc}
We have investigated feedback control of cold atomic ensembles in particular for two-component spin coherent states. Measuring the environment coupled to the system with the operator $S^z$, we can make an estimate of the entire state $\rho_c$ of the BEC including estimates of $ \langle s^{x,y,z} \rangle$ based on the measurement results by the quantum filtering theorem. The  analysis of convergence to the target state is based on the application of  LaSalle's invariance theorem, which allows us to asymptotically prepares a particular quantum state in the sense that   $\langle X \rangle_t= \mathrm{Tr} (\rho_c X )$   as
$t\rightarrow +\infty$ for all $X$.  We have combined the above two theorems  to design a control law to drive the BEC state to an arbitrary state on the
Bloch sphere.  This yields the control laws as summarized in Table \ref{tab1} to control each of the $ S^{x,y,z} $ directions.  While it is possible to find what control law should be implemented for a particular state, a more difficult task is to know in advance the particular parameters, which would involve solving the conditional master equation at steady-state.  In practice however, finding the parameters for the control law does not involve a large search space, and hence can be found efficiently using standard search techniques.

 Our method
separates the control problem into an estimation step (filtering) and a control step
based on the estimates only. The simulation results show that the feedback control works very well to prepare an arbitrary spin coherent state by using various combinations of the feedback signal and control Hamiltonian, which renders the filter stochastically stable around the target state. We expect that this technique should be of particular relevance for advanced control methods for performing generalized error correction type operations such as that performed in Ref. \cite{vanderbruggen2013}. This would allow a method of storing quantum information in a desired target state virtually indefinitely by counteracting the effects of decoherence.

\section*{ACKNOWLEDGMENTS}
We thank B. Qi, S. S. Szigeti, M. R. Hush, and R. Johansson for illuminating discussions.
This work is supported by the Shanghai Research Challenge Fund, New York University Global Seed Grants for Collaborative Research, National Natural Science Foundation of China grant 61571301, and the Thousand Talents Program for Distinguished Young Scholars.

\appendix

\section{Derivation of conditional master equation for ensembles}

Here we rederive the conditional master equation  (\ref{conditional-master1})  for the case of ensembles.  Consider an ensemble of $ N $ atoms labeled by an index $ l $, each having two ground states  $ |g 1\rangle_l , | g 2\rangle_l $ which serve as the logical states.  Each ground state state has an excited state which connected by a photon of suitable polarization and are labeled $ |e 1\rangle_l , | e 2\rangle_l $.  The energy of the states may be written
\begin{align}
\hat{H}_{\text S} = \sum_{l=1}^N \sum_{n=g,e} \sum_{j=1,2} \omega_{nj} | n j \rangle_l \langle n j |_l .
\end{align}
Illumination with an off-resonant electromagnetic field and adiabatic elimination of the excited states gives an effective Hamiltonian
\begin{align}
\hat{H}_{\text eff} = \sum_{l=1}^N G_1 \left( | g 1 \rangle_l \langle g 1 |_l + G_2  | g 2 \rangle_l \langle g 2 |_l  \right) ,
\end{align}
where
\begin{align}
G_j = -2| E^-|^2 \frac{| \langle g j |_l \bm{d}_j \cdot \varepsilon | e j \rangle_l |^2 }{\Delta_j}
\end{align}
and $  \varepsilon $ is the polarization vector. Here we have assumed that the strength of the electromagnetic field is homogenous for all the atoms in the ensemble, such that the $ l $-dependence drops out.  Now that the excited states have been eliminated, we henceforth drop the ground state labels and write $ | g j \rangle_l  \rightarrow |  j \rangle_l $. It is convenient to introduce operators
\begin{align}
n_j \rightarrow \sum_{l=1}^N |j \rangle_l \langle j |_l,
\end{align}
such that the total spin operators are
\begin{align}
S^x & = \sum_{l=1}^N \left( |1\rangle_l \langle 2 |_l + |2\rangle_l \langle 1 |_l \right) \nonumber \\
S^y & = \sum_{l=1}^N \left( -i|1\rangle_l \langle 2 |_l +i |2\rangle_l \langle 1 |_l \right) \nonumber \\
S^z & = \sum_{l=1}^N \left( |1\rangle_l \langle 1 |_l - |2\rangle_l \langle 2 |_l \right) \nonumber \\
N & = \sum_{l=1}^N \left( |1\rangle_l \langle 1 |_l + |2\rangle_l \langle 2 |_l \right)  .
\end{align}
The effective Hamiltonian can thus be written in the same form as (\ref{spin-form-H}).  The measurement operator
then takes the same form as (\ref{measop}) but with
\begin{align}
M_j(\emph{\textbf{x}})=\chi_j(\emph{\textbf{x}}) .
\end{align}
The remaining steps from (\ref{dexpr}) to (\ref{conditional-master1}) then follow in the same way but with a redefinition of the spin operators as above.



\begin{thebibliography}{43}
\expandafter\ifx\csname natexlab\endcsname\relax\def\natexlab#1{#1}\fi
\expandafter\ifx\csname bibnamefont\endcsname\relax
  \def\bibnamefont#1{#1}\fi
\expandafter\ifx\csname bibfnamefont\endcsname\relax
  \def\bibfnamefont#1{#1}\fi
\expandafter\ifx\csname citenamefont\endcsname\relax
  \def\citenamefont#1{#1}\fi
\expandafter\ifx\csname url\endcsname\relax
  \def\url#1{\texttt{#1}}\fi
\expandafter\ifx\csname urlprefix\endcsname\relax\def\urlprefix{URL }\fi
\providecommand{\bibinfo}[2]{#2}
\providecommand{\eprint}[2][]{\url{#2}}

\bibitem[{\citenamefont{Treutlein et~al.}(2006)\citenamefont{Treutlein,
  Steinmetz, Colombe, Lev, Hommelhoff, Reichel, Greiner, Mandel, Widera, Rom
  et~al.}}]{treutlein2006}
\bibinfo{author}{\bibfnamefont{P.}~\bibnamefont{Treutlein}},
  \bibinfo{author}{\bibfnamefont{T.}~\bibnamefont{Steinmetz}},
  \bibinfo{author}{\bibfnamefont{Y.}~\bibnamefont{Colombe}},
  \bibinfo{author}{\bibfnamefont{B.}~\bibnamefont{Lev}},
  \bibinfo{author}{\bibfnamefont{P.}~\bibnamefont{Hommelhoff}},
  \bibinfo{author}{\bibfnamefont{J.}~\bibnamefont{Reichel}},
  \bibinfo{author}{\bibfnamefont{M.}~\bibnamefont{Greiner}},
  \bibinfo{author}{\bibfnamefont{O.}~\bibnamefont{Mandel}},
  \bibinfo{author}{\bibfnamefont{A.}~\bibnamefont{Widera}},
  \bibinfo{author}{\bibfnamefont{T.}~\bibnamefont{Rom}}, \bibnamefont{et~al.},
  \bibinfo{journal}{Fortschr. Phys.} \textbf{\bibinfo{volume}{54}},
  \bibinfo{pages}{702} (\bibinfo{year}{2006}).

\bibitem[{\citenamefont{B{\"o}hi et~al.}(2009)\citenamefont{B{\"o}hi, Riedel,
  Hoffrogge, Reichel, H{\"a}nsch, and Treutlein}}]{bohi2009}
\bibinfo{author}{\bibfnamefont{P.}~\bibnamefont{B{\"o}hi}},
  \bibinfo{author}{\bibfnamefont{M.~F.} \bibnamefont{Riedel}},
  \bibinfo{author}{\bibfnamefont{J.}~\bibnamefont{Hoffrogge}},
  \bibinfo{author}{\bibfnamefont{J.}~\bibnamefont{Reichel}},
  \bibinfo{author}{\bibfnamefont{T.}~\bibnamefont{H{\"a}nsch}},
  \bibnamefont{and}
  \bibinfo{author}{\bibfnamefont{P.}~\bibnamefont{Treutlein}},
  \bibinfo{journal}{Nature Physics} \textbf{\bibinfo{volume}{5}},
  \bibinfo{pages}{592} (\bibinfo{year}{2009}).

\bibitem[{\citenamefont{Riedel et~al.}(2010)\citenamefont{Riedel, B{\"o}hi, Li,
  H{\"a}nsch, Sinatra, and Treutlein}}]{riedel2010}
\bibinfo{author}{\bibfnamefont{M.~F.} \bibnamefont{Riedel}},
  \bibinfo{author}{\bibfnamefont{P.}~\bibnamefont{B{\"o}hi}},
  \bibinfo{author}{\bibfnamefont{Y.}~\bibnamefont{Li}},
  \bibinfo{author}{\bibfnamefont{T.~W.} \bibnamefont{H{\"a}nsch}},
  \bibinfo{author}{\bibfnamefont{A.}~\bibnamefont{Sinatra}}, \bibnamefont{and}
  \bibinfo{author}{\bibfnamefont{P.}~\bibnamefont{Treutlein}},
  \bibinfo{journal}{Nature} \textbf{\bibinfo{volume}{464}},
  \bibinfo{pages}{1170} (\bibinfo{year}{2010}).

\bibitem[{\citenamefont{Gross}(2012)}]{gross2012}
\bibinfo{author}{\bibfnamefont{C.}~\bibnamefont{Gross}}, \bibinfo{journal}{J.
  Phys. B: At. Mol. Opt. Phys.} \textbf{\bibinfo{volume}{45}},
  \bibinfo{pages}{103001} (\bibinfo{year}{2012}).

\bibitem[{\citenamefont{Buluta and Nori}(2009)}]{buluta2009}
\bibinfo{author}{\bibfnamefont{I.}~\bibnamefont{Buluta}} \bibnamefont{and}
  \bibinfo{author}{\bibfnamefont{F.}~\bibnamefont{Nori}},
  \bibinfo{journal}{Science} \textbf{\bibinfo{volume}{326}},
  \bibinfo{pages}{108} (\bibinfo{year}{2009}).

\bibitem[{\citenamefont{Bloch}(2005)}]{bloch2005}
\bibinfo{author}{\bibfnamefont{I.}~\bibnamefont{Bloch}},
  \bibinfo{journal}{Nature Physics} \textbf{\bibinfo{volume}{1}},
  \bibinfo{pages}{23} (\bibinfo{year}{2005}).

\bibitem[{\citenamefont{Ladd et~al.}(2010)\citenamefont{Ladd, Jelezko,
  Laflamme, Nakamura, Monroe, and O’Brien}}]{ladd2010}
\bibinfo{author}{\bibfnamefont{T.~D.} \bibnamefont{Ladd}},
  \bibinfo{author}{\bibfnamefont{F.}~\bibnamefont{Jelezko}},
  \bibinfo{author}{\bibfnamefont{R.}~\bibnamefont{Laflamme}},
  \bibinfo{author}{\bibfnamefont{Y.}~\bibnamefont{Nakamura}},
  \bibinfo{author}{\bibfnamefont{C.}~\bibnamefont{Monroe}}, \bibnamefont{and}
  \bibinfo{author}{\bibfnamefont{J.~L.} \bibnamefont{O’Brien}},
  \bibinfo{journal}{Nature} \textbf{\bibinfo{volume}{464}}, \bibinfo{pages}{45}
  (\bibinfo{year}{2010}).

\bibitem[{\citenamefont{Brion et~al.}(2007)\citenamefont{Brion, Molmer, and
  Saffman}}]{brion2007}
\bibinfo{author}{\bibfnamefont{E.}~\bibnamefont{Brion}},
  \bibinfo{author}{\bibfnamefont{K.}~\bibnamefont{Molmer}}, \bibnamefont{and}
  \bibinfo{author}{\bibfnamefont{M.}~\bibnamefont{Saffman}},
  \bibinfo{journal}{Phys. Rev. Lett.} \textbf{\bibinfo{volume}{99}},
  \bibinfo{pages}{260501} (\bibinfo{year}{2007}).

\bibitem[{\citenamefont{Lukin et~al.}(2001)\citenamefont{Lukin, Fleischhauer,
  Cote, Duan, Jaksch, Cirac, and Zoller}}]{lukin2001}
\bibinfo{author}{\bibfnamefont{M.~D.} \bibnamefont{Lukin}},
  \bibinfo{author}{\bibfnamefont{M.}~\bibnamefont{Fleischhauer}},
  \bibinfo{author}{\bibfnamefont{R.}~\bibnamefont{Cote}},
  \bibinfo{author}{\bibfnamefont{L.~M.} \bibnamefont{Duan}},
  \bibinfo{author}{\bibfnamefont{D.}~\bibnamefont{Jaksch}},
  \bibinfo{author}{\bibfnamefont{J.~I.} \bibnamefont{Cirac}}, \bibnamefont{and}
  \bibinfo{author}{\bibfnamefont{P.}~\bibnamefont{Zoller}},
  \bibinfo{journal}{Phys. Rev. Lett.} \textbf{\bibinfo{volume}{87}},
  \bibinfo{pages}{037901} (\bibinfo{year}{2001}).

\bibitem[{\citenamefont{Bao et~al.}(2012)\citenamefont{Bao, Xu, Li, Yuan, Lu,
  and Pan}}]{bao2012}
\bibinfo{author}{\bibfnamefont{X.-H.} \bibnamefont{Bao}},
  \bibinfo{author}{\bibfnamefont{X.-F.} \bibnamefont{Xu}},
  \bibinfo{author}{\bibfnamefont{C.-M.} \bibnamefont{Li}},
  \bibinfo{author}{\bibfnamefont{Z.-S.} \bibnamefont{Yuan}},
  \bibinfo{author}{\bibfnamefont{C.-Y.} \bibnamefont{Lu}}, \bibnamefont{and}
  \bibinfo{author}{\bibfnamefont{J.-W.} \bibnamefont{Pan}},
  \bibinfo{journal}{Proc. Nat. Acad. Sci.} \textbf{\bibinfo{volume}{109}},
  \bibinfo{pages}{20347} (\bibinfo{year}{2012}).

\bibitem[{\citenamefont{Julsgaard et~al.}(2001)\citenamefont{Julsgaard,
  Kozhekin, and Polzik}}]{julsgaard2001}
\bibinfo{author}{\bibfnamefont{B.}~\bibnamefont{Julsgaard}},
  \bibinfo{author}{\bibfnamefont{A.}~\bibnamefont{Kozhekin}}, \bibnamefont{and}
  \bibinfo{author}{\bibfnamefont{E.}~\bibnamefont{Polzik}},
  \bibinfo{journal}{Nature} \textbf{\bibinfo{volume}{413}},
  \bibinfo{pages}{400} (\bibinfo{year}{2001}).

\bibitem[{\citenamefont{Krauter et~al.}(2012)\citenamefont{Krauter, Salart,
  Muschik, Petersen, Shen, Fernholz, and Polzik}}]{krauter2012}
\bibinfo{author}{\bibfnamefont{H.}~\bibnamefont{Krauter}},
  \bibinfo{author}{\bibfnamefont{D.}~\bibnamefont{Salart}},
  \bibinfo{author}{\bibfnamefont{C.~A.} \bibnamefont{Muschik}},
  \bibinfo{author}{\bibfnamefont{J.~M.} \bibnamefont{Petersen}},
  \bibinfo{author}{\bibfnamefont{H.}~\bibnamefont{Shen}},
  \bibinfo{author}{\bibfnamefont{T.}~\bibnamefont{Fernholz}}, \bibnamefont{and}
  \bibinfo{author}{\bibfnamefont{E.~S.} \bibnamefont{Polzik}},
  \bibinfo{journal}{Nature Phys.} \textbf{\bibinfo{volume}{9}},
  \bibinfo{pages}{400} (\bibinfo{year}{2012}).

\bibitem[{\citenamefont{Braunstein and van Loock}(2005)}]{braunstein2005}
\bibinfo{author}{\bibfnamefont{S.}~\bibnamefont{Braunstein}} \bibnamefont{and}
  \bibinfo{author}{\bibfnamefont{P.}~\bibnamefont{van Loock}},
  \bibinfo{journal}{Rev. Mod. Phys.} \textbf{\bibinfo{volume}{77}},
  \bibinfo{pages}{513} (\bibinfo{year}{2005}).

\bibitem[{\citenamefont{Byrnes et~al.}(2012)\citenamefont{Byrnes, Wen, and
  Yamamoto}}]{byrnes2012}
\bibinfo{author}{\bibfnamefont{T.}~\bibnamefont{Byrnes}},
  \bibinfo{author}{\bibfnamefont{K.}~\bibnamefont{Wen}}, \bibnamefont{and}
  \bibinfo{author}{\bibfnamefont{Y.}~\bibnamefont{Yamamoto}},
  \bibinfo{journal}{Phys. Rev. A} \textbf{\bibinfo{volume}{85}},
  \bibinfo{pages}{040306} (\bibinfo{year}{2012}).

\bibitem[{\citenamefont{Byrnes et~al.}(2015)\citenamefont{Byrnes, Rosseau,
  Khosla, Pyrkov, Thomasen, Mukai, Koyama, Abdelrahman, and
  Ilo-Okeke}}]{byrnes2014}
\bibinfo{author}{\bibfnamefont{T.}~\bibnamefont{Byrnes}},
  \bibinfo{author}{\bibfnamefont{D.}~\bibnamefont{Rosseau}},
  \bibinfo{author}{\bibfnamefont{M.}~\bibnamefont{Khosla}},
  \bibinfo{author}{\bibfnamefont{A.}~\bibnamefont{Pyrkov}},
  \bibinfo{author}{\bibfnamefont{A.}~\bibnamefont{Thomasen}},
  \bibinfo{author}{\bibfnamefont{T.}~\bibnamefont{Mukai}},
  \bibinfo{author}{\bibfnamefont{S.}~\bibnamefont{Koyama}},
  \bibinfo{author}{\bibfnamefont{A.}~\bibnamefont{Abdelrahman}},
  \bibnamefont{and}
  \bibinfo{author}{\bibfnamefont{E.}~\bibnamefont{Ilo-Okeke}},
  \bibinfo{journal}{Opt. Comm.} \textbf{\bibinfo{volume}{337}},
  \bibinfo{pages}{102} (\bibinfo{year}{2015}).

\bibitem[{\citenamefont{Pyrkov and Byrnes}(2013)}]{pyrkov2013}
\bibinfo{author}{\bibfnamefont{A.}~\bibnamefont{Pyrkov}} \bibnamefont{and}
  \bibinfo{author}{\bibfnamefont{T.}~\bibnamefont{Byrnes}},
  \bibinfo{journal}{New J. Phys..} \textbf{\bibinfo{volume}{15}},
  \bibinfo{pages}{093019} (\bibinfo{year}{2013}).

\bibitem[{\citenamefont{Anderson et~al.}(1995)\citenamefont{Anderson, Ensher,
  Matthews, Wieman, and Cornell}}]{anderson1995}
\bibinfo{author}{\bibfnamefont{M.~H.} \bibnamefont{Anderson}},
  \bibinfo{author}{\bibfnamefont{J.~R.} \bibnamefont{Ensher}},
  \bibinfo{author}{\bibfnamefont{M.~R.} \bibnamefont{Matthews}},
  \bibinfo{author}{\bibfnamefont{C.~E.} \bibnamefont{Wieman}},
  \bibnamefont{and} \bibinfo{author}{\bibfnamefont{E.~A.}
  \bibnamefont{Cornell}}, \bibinfo{journal}{Science}
  \textbf{\bibinfo{volume}{269}}, \bibinfo{pages}{198} (\bibinfo{year}{1995}).

\bibitem[{\citenamefont{Andrews et~al.}(1997)\citenamefont{Andrews, Kurn,
  Miesner, Durfee, Townsend, Inouye, and Ketterle}}]{andrews1997}
\bibinfo{author}{\bibfnamefont{M.~R.} \bibnamefont{Andrews}},
  \bibinfo{author}{\bibfnamefont{D.~M.} \bibnamefont{Kurn}},
  \bibinfo{author}{\bibfnamefont{H.-J.} \bibnamefont{Miesner}},
  \bibinfo{author}{\bibfnamefont{D.~S.} \bibnamefont{Durfee}},
  \bibinfo{author}{\bibfnamefont{C.~G.} \bibnamefont{Townsend}},
  \bibinfo{author}{\bibfnamefont{S.}~\bibnamefont{Inouye}}, \bibnamefont{and}
  \bibinfo{author}{\bibfnamefont{W.}~\bibnamefont{Ketterle}},
  \bibinfo{journal}{Phys. Rev. Lett.} \textbf{\bibinfo{volume}{79}},
  \bibinfo{pages}{553} (\bibinfo{year}{1997}).

\bibitem[{\citenamefont{Vestergaard-Hau
  et~al.}(1998)\citenamefont{Vestergaard-Hau, Busch, Liu, Burns, and
  Golovchenko}}]{vestergaardhau1998}
\bibinfo{author}{\bibfnamefont{L.V.}~\bibnamefont{Hau}},
  \bibinfo{author}{\bibfnamefont{B.~D.} \bibnamefont{Busch}},
  \bibinfo{author}{\bibfnamefont{C.}~\bibnamefont{Liu}},
  \bibinfo{author}{\bibfnamefont{Z.}~\bibnamefont{Dutton}},
  \bibinfo{author}{\bibfnamefont{M.~M.} \bibnamefont{Burns}}, \bibnamefont{and}
  \bibinfo{author}{\bibfnamefont{J.~A.} \bibnamefont{Golovchenko}},
  \bibinfo{journal}{Phys. Rev. A} \textbf{\bibinfo{volume}{58}},
  \bibinfo{pages}{54(R)} (\bibinfo{year}{1998}).

\bibitem[{\citenamefont{DePue et~al.}(2000)\citenamefont{DePue, Winoto, Han,
  and Weiss}}]{depue2000}
\bibinfo{author}{\bibfnamefont{M.~T.} \bibnamefont{DePue}},
  \bibinfo{author}{\bibfnamefont{S.~L.} \bibnamefont{Winoto}},
  \bibinfo{author}{\bibfnamefont{D.~J.} \bibnamefont{Han}}, \bibnamefont{and}
  \bibinfo{author}{\bibfnamefont{D.~S.} \bibnamefont{Weiss}},
  \bibinfo{journal}{Optics Communications} \textbf{\bibinfo{volume}{180}},
  \bibinfo{pages}{73} (\bibinfo{year}{2000}).

\bibitem[{\citenamefont{Bradley et~al.}(1997)\citenamefont{Bradley, Sackett,
  and Hulet}}]{bradley1997}
\bibinfo{author}{\bibfnamefont{C.~C.} \bibnamefont{Bradley}},
  \bibinfo{author}{\bibfnamefont{C.~A.} \bibnamefont{Sackett}},
  \bibnamefont{and} \bibinfo{author}{\bibfnamefont{R.~G.} \bibnamefont{Hulet}},
  \bibinfo{journal}{Phys. Rev. Lett.} \textbf{\bibinfo{volume}{78}},
  \bibinfo{pages}{985} (\bibinfo{year}{1997}).

\bibitem[{\citenamefont{Ilo-Okeke and
  Byrnes}(2014{\natexlab{a}})}]{ilookeke2014}
\bibinfo{author}{\bibfnamefont{E.~O.} \bibnamefont{Ilo-Okeke}}
  \bibnamefont{and} \bibinfo{author}{\bibfnamefont{T.}~\bibnamefont{Byrnes}},
  \bibinfo{journal}{Phys. Rev. Lett.} \textbf{\bibinfo{volume}{112}},
  \bibinfo{pages}{233602} (\bibinfo{year}{2014}{\natexlab{a}}).

\bibitem[{\citenamefont{Vanderbruggen et~al.}(2013)\citenamefont{Vanderbruggen,
  Kohlhaas, Bertoldi, Bernon, Aspect, Landragin, and
  Bouyer}}]{vanderbruggen2013}
\bibinfo{author}{\bibfnamefont{T.}~\bibnamefont{Vanderbruggen}},
  \bibinfo{author}{\bibfnamefont{R.}~\bibnamefont{Kohlhaas}},
  \bibinfo{author}{\bibfnamefont{A.}~\bibnamefont{Bertoldi}},
  \bibinfo{author}{\bibfnamefont{S.}~\bibnamefont{Bernon}},
  \bibinfo{author}{\bibfnamefont{A.}~\bibnamefont{Aspect}},
  \bibinfo{author}{\bibfnamefont{A.}~\bibnamefont{Landragin}},
  \bibnamefont{and} \bibinfo{author}{\bibfnamefont{P.}~\bibnamefont{Bouyer}},
  \bibinfo{journal}{Phys. Rev. Lett.} \textbf{\bibinfo{volume}{110}},
  \bibinfo{pages}{210503} (\bibinfo{year}{2013}).

\bibitem[{\citenamefont{Shiga and Takeuchi}(2013)}]{shiga2012}
\bibinfo{author}{\bibfnamefont{N.}~\bibnamefont{Shiga}} \bibnamefont{and}
  \bibinfo{author}{\bibfnamefont{M.}~\bibnamefont{Takeuchi}},
  \bibinfo{journal}{New J. Phys.} \textbf{\bibinfo{volume}{14}},
  \bibinfo{pages}{023034} (\bibinfo{year}{2013}).

\bibitem[{\citenamefont{Wiseman and Milburn}(1993)}]{WM93}
\bibinfo{author}{\bibfnamefont{H.~M.} \bibnamefont{Wiseman}} \bibnamefont{and}
  \bibinfo{author}{\bibfnamefont{G.~J.} \bibnamefont{Milburn}},
  \bibinfo{journal}{Phys. Rev. Lett.} \textbf{\bibinfo{volume}{70}},
  \bibinfo{pages}{548} (\bibinfo{year}{1993}).

\bibitem[{\citenamefont{Wiseman}(1994)}]{W94}
\bibinfo{author}{\bibfnamefont{H.~M.} \bibnamefont{Wiseman}},
  \bibinfo{journal}{Phys. Rev. A} \textbf{\bibinfo{volume}{49}},
  \bibinfo{pages}{2133} (\bibinfo{year}{1994}).

\bibitem[{\citenamefont{Doherty and Jacobs}(1999)}]{DJ99}
\bibinfo{author}{\bibfnamefont{A.~C.} \bibnamefont{Doherty}} \bibnamefont{and}
  \bibinfo{author}{\bibfnamefont{K.}~\bibnamefont{Jacobs}},
  \bibinfo{journal}{Phys. Rev. A} \textbf{\bibinfo{volume}{60}},
  \bibinfo{pages}{2700} (\bibinfo{year}{1999}).

\bibitem[{\citenamefont{Doherty et~al.}(2000)\citenamefont{Doherty, Habib,
  Jacobs, Mabuchi, and Tan}}]{DHJMT2000}
\bibinfo{author}{\bibfnamefont{A.~C.} \bibnamefont{Doherty}},
  \bibinfo{author}{\bibfnamefont{S.}~\bibnamefont{Habib}},
  \bibinfo{author}{\bibfnamefont{K.}~\bibnamefont{Jacobs}},
  \bibinfo{author}{\bibfnamefont{H.}~\bibnamefont{Mabuchi}}, \bibnamefont{and}
  \bibinfo{author}{\bibfnamefont{S.~M.} \bibnamefont{Tan}},
  \bibinfo{journal}{Phys. Rev. A} \textbf{\bibinfo{volume}{62}},
  \bibinfo{pages}{012105} (\bibinfo{year}{2000}).

\bibitem[{\citenamefont{Wiseman and Doherty}(2005)}]{WD2005}
\bibinfo{author}{\bibfnamefont{H.~M.} \bibnamefont{Wiseman}} \bibnamefont{and}
  \bibinfo{author}{\bibfnamefont{A.~C.} \bibnamefont{Doherty}},
  \bibinfo{journal}{Phys. Rev. Lett.} \textbf{\bibinfo{volume}{94}},
  \bibinfo{pages}{070405} (\bibinfo{year}{2005}).

\bibitem[{\citenamefont{Wiseman and Milburn}(2009)}]{WM2009}
\bibinfo{author}{\bibfnamefont{H.~W.} \bibnamefont{Wiseman}} \bibnamefont{and}
  \bibinfo{author}{\bibfnamefont{G.~J.} \bibnamefont{Milburn}},
  \emph{\bibinfo{title}{Quantum Measurement and Control}}
  (\bibinfo{publisher}{Cambridge University Press},
  \bibinfo{address}{Cambridge, UK}, \bibinfo{year}{2009}).

\bibitem[{\citenamefont{Chu}(2002)}]{CHUS02}
\bibinfo{author}{\bibfnamefont{S.}~\bibnamefont{Chu}},
  \bibinfo{journal}{Nature} \textbf{\bibinfo{volume}{416}},
  \bibinfo{pages}{206} (\bibinfo{year}{2002}).

\bibitem[{\citenamefont{Joe-Wong et~al.}(2013)\citenamefont{Joe-Wong, Ho, Long,
  Rabitz, and Wu}}]{WHLRW13}
\bibinfo{author}{\bibfnamefont{C.}~\bibnamefont{Joe-Wong}},
  \bibinfo{author}{\bibfnamefont{T.}~\bibnamefont{Ho}},
  \bibinfo{author}{\bibfnamefont{R.}~\bibnamefont{Long}},
  \bibinfo{author}{\bibfnamefont{H.}~\bibnamefont{Rabitz}}, \bibnamefont{and}
  \bibinfo{author}{\bibfnamefont{R.}~\bibnamefont{Wu}}, \bibinfo{journal}{The
  Journal of Chemical Physics} \textbf{\bibinfo{volume}{138}},
  \bibinfo{eid}{124114} (\bibinfo{year}{2013}).

\bibitem[{\citenamefont{Wiseman and Bouten}(2008)}]{WB08}
\bibinfo{author}{\bibfnamefont{H.~M.} \bibnamefont{Wiseman}} \bibnamefont{and}
  \bibinfo{author}{\bibfnamefont{L.}~\bibnamefont{Bouten}},
  \bibinfo{journal}{Quantum Information Processing}
  \textbf{\bibinfo{volume}{7}}, \bibinfo{pages}{71} (\bibinfo{year}{2008}).

\bibitem[{\citenamefont{Hou et~al.}(2012)\citenamefont{Hou, Khan, Yi, Dong, and
  Petersen}}]{HKYDI2012}
\bibinfo{author}{\bibfnamefont{S.~C.} \bibnamefont{Hou}},
  \bibinfo{author}{\bibfnamefont{M.~A.} \bibnamefont{Khan}},
  \bibinfo{author}{\bibfnamefont{X.~X.} \bibnamefont{Yi}},
  \bibinfo{author}{\bibfnamefont{D.}~\bibnamefont{Dong}}, \bibnamefont{and}
  \bibinfo{author}{\bibfnamefont{I.~R.} \bibnamefont{Petersen}},
  \bibinfo{journal}{Phys. Rev. A} \textbf{\bibinfo{volume}{86}},
  \bibinfo{pages}{022321} (\bibinfo{year}{2012}).

\bibitem[{\citenamefont{Szigeti et~al.}(2009)\citenamefont{Szigeti, Hush,
  Carvalho, and Hope}}]{szigeti2009}
\bibinfo{author}{\bibfnamefont{S.~S.} \bibnamefont{Szigeti}},
  \bibinfo{author}{\bibfnamefont{M.~R.} \bibnamefont{Hush}},
  \bibinfo{author}{\bibfnamefont{A.~R.~R.} \bibnamefont{Carvalho}}, \bibnamefont{and}
   \bibinfo{author}{\bibfnamefont{J.~J.} \bibnamefont{Hope}},
  \bibinfo{journal}{Phys. Rev. A} \textbf{\bibinfo{volume}{80}},
  \bibinfo{pages}{013614} (\bibinfo{year}{2009}).

\bibitem[{\citenamefont{Szigeti et~al.}(2010)\citenamefont{Szigeti, Hush,
  Carvalho, and Hope}}]{szigeti2010}
\bibinfo{author}{\bibfnamefont{S.~S.} \bibnamefont{Szigeti}},
  \bibinfo{author}{\bibfnamefont{M.~R.} \bibnamefont{Hush}},
  \bibinfo{author}{\bibfnamefont{A.~R.~R.} \bibnamefont{Carvalho}}, \bibnamefont{and}
   \bibinfo{author}{\bibfnamefont{J.~J.} \bibnamefont{Hope}},
  \bibinfo{journal}{Phys. Rev. A} \textbf{\bibinfo{volume}{82}},
  \bibinfo{pages}{043632} (\bibinfo{year}{2010}).

\bibitem[{\citenamefont{LaSalle and Lefschetz}(1961)}]{LL1961}
\bibinfo{author}{\bibfnamefont{J.}~\bibnamefont{LaSalle}} \bibnamefont{and}
  \bibinfo{author}{\bibfnamefont{S.}~\bibnamefont{Lefschetz}},
  \emph{\bibinfo{title}{Stability by Lyapunovs direct method with
  applications}} (\bibinfo{publisher}{Academic Press}, \bibinfo{address}{New
  York}, \bibinfo{year}{1961}).

\bibitem[{\citenamefont{Wang and Schirmer}(2010)}]{wang2010}
\bibinfo{author}{\bibfnamefont{X.}~\bibnamefont{Wang}} \bibnamefont{and}
  \bibinfo{author}{\bibfnamefont{S.~G.} \bibnamefont{Schirmer}},
  \bibinfo{journal}{IEEE Transactions on Automatic Control}
  \textbf{\bibinfo{volume}{55}}, \bibinfo{pages}{2259} (\bibinfo{year}{2010}).

\bibitem[{\citenamefont{Ilo-Okeke and Byrnes}(2014{\natexlab{b}})}]{ilookeke14}
\bibinfo{author}{\bibfnamefont{E.~O.} \bibnamefont{Ilo-Okeke}}
  \bibnamefont{and} \bibinfo{author}{\bibfnamefont{T.}~\bibnamefont{Byrnes}},
  \bibinfo{journal}{Phys. Rev. Lett.} \textbf{\bibinfo{volume}{112}},
  \bibinfo{pages}{233602} (\bibinfo{year}{2014}{\natexlab{b}}).

\bibitem[{\citenamefont{Massar and Popescu}(1995)}]{massar1995}
\bibinfo{author}{\bibfnamefont{S.}~\bibnamefont{Massar}} \bibnamefont{and}
  \bibinfo{author}{\bibfnamefont{S.}~\bibnamefont{Popescu}},
  \bibinfo{journal}{Phys. Rev. Lett.} \textbf{\bibinfo{volume}{74}},
  \bibinfo{pages}{1259} (\bibinfo{year}{1995}).



\bibitem[{\citenamefont{Ilo-Okeke and
  Byrnes}(2016)}]{ilookeke2016}
\bibinfo{author}{\bibfnamefont{E.~O.} \bibnamefont{Ilo-Okeke}}
  \bibnamefont{and} \bibinfo{author}{\bibfnamefont{T.}~\bibnamefont{Byrnes}},
  \bibinfo{journal}{Phys. Rev. A} \textbf{\bibinfo{volume}{94}},
  \bibinfo{pages}{013617} (\bibinfo{year}{2016}).

\bibitem[{\citenamefont{Gardiner and Zoller}(2004)}]{GZ04}
\bibinfo{author}{\bibfnamefont{G.}~\bibnamefont{Gardiner}} \bibnamefont{and}
  \bibinfo{author}{\bibfnamefont{P.}~\bibnamefont{Zoller}},
  \emph{\bibinfo{title}{Quantum Noise}} (\bibinfo{publisher}{Springer},
  \bibinfo{address}{Berlin}, \bibinfo{year}{2004}),
  chap.~\bibinfo{chapter}{11}.

\bibitem[{\citenamefont{Belavkin}(1994)}]{BVP94}
\bibinfo{author}{\bibfnamefont{V.~P.} \bibnamefont{Belavkin}},
  \bibinfo{journal}{SIAM J. Control Optim.} \textbf{\bibinfo{volume}{24}},
  \bibinfo{pages}{685} (\bibinfo{year}{1994}).

\bibitem[{\citenamefont{L.~Bouten and James}(2007)}]{BHJ2007}
\bibinfo{author}{\bibfnamefont{R. van Handel}, \bibnamefont{L.~Bouten}},
  \bibnamefont{and} \bibinfo{author}{\bibfnamefont{M.~R.} \bibnamefont{James}},
  \bibinfo{journal}{SIAM J. Control Optim.} \textbf{\bibinfo{volume}{46}},
  \bibinfo{pages}{2199} (\bibinfo{year}{2007}).
\end{thebibliography}

\end{document}